\documentclass[twocolumn,showpacs,preprintnumbers,amsmath,amssymb]{revtex4}

\usepackage{graphicx}
\usepackage {amsmath}
\usepackage{hyperref}

\begin{document}

\title{Stable and unstable regimes in Bose-Fermi mixture with attraction
between components}

\author{A.M. Belemuk}
\affiliation{Institute for High Pressure Physics, Russian Academy
of Sciences, Troitsk 142190, Moscow Region, Russia}

\author{S.-T. Chui}
\affiliation{Bartol Research Institute, University of Delaware,
Newark, DE 19716}

\author{V.N. Ryzhov}
\affiliation{Institute for High Pressure Physics, Russian Academy
of Sciences, Troitsk 142190, Moscow Region, Russia}

\date{\today}

\begin{abstract}
A collapse of the trapped boson- fermion mixture with the
attraction between bosons and fermions is investigated in the
framework of the effective Hamiltonian for the Bose system. The
properties of the $^{87}$Rb and $^{40}$K mixture are analyzed
quantitatively at $T= 0$. We find numerically solutions of
modified Gross- Pitaevskii equation which continuously go from
stable to unstable branch. We discuss the relation of the onset of
collapse with macroscopic properties of the system. A comparison
with the case of a Bose condensate of atomic $^7Li$ system is
given.
\end{abstract}


\maketitle

\section{Introduction}

Bose-Einstein condensation (BEC) in ultracold atomic gas clouds
with repulsive and attractive interatomic interactions
\cite{[1],[2],[3]} have been the subject of intense theoretical
and experimental interest in recent years. Besides the studies
using the bosonic atoms, growing interest is focused on the
cooling of fermionic atoms \cite{sci99}. Cooling of trapped
fermionic atoms to a temperature regime where a Fermi gas can be
considered as degenerate has been possible by sympathetic cooling
in the presence of a second boson or fermion component. Quantum
degeneracy was first reached with mixtures of bosonic $^7$Li and
fermionic $^6$Li atoms \cite{sci01,prl1}. Later, experiments to
cool mixtures of $^{23}$Na and $^6$Li \cite{NaLi}, as well as
$^{87}$Rb and $^{40}$K \cite{sci02bf, KRb}, to ultralow
temperatures succeeded. The Bose gas, which can be cooled
evaporatively, is used as a coolant, the fermionic system being in
thermal equilibrium with the cold Bose gas through boson-fermion
interaction in the region of overlapping of the systems.

Collisional interaction between bosons and fermions greatly affect
the properties of the degenerate mixture. Theoretical
considerations predict the phenomenon of component separation for
systems with a positive coupling constant \cite{Nygaard99},
instabilities and significant modification of the properties of
individual component in a case of boson-fermion attraction
\cite{Roth02, Molmer98}. Effect of density fluctuations in a Bose
condensate on the fermion-fermion interaction with relevant
implications for the achievement of fermionic superfluidity has
been investigated in \cite{stoof2}. The presence of a sufficiently
attractive boson-fermion interaction can bring about the formation
of stable fermionic bright solitons \cite{Adhikari05}.

The simultaneous collapse of the two species has been observed in
a $^{40}K-^{87}Rb$ mixture by Modugno and co-workers
\cite{sci02bf}. K and Rb atoms were prepared in doubly polarized
states $|F=9/2, m_F=9/2>$ and $|2,2>$, respectively. As was shown,
as the number of bosons is increased there is an instability value
$N_{Bc}$ at which a discontinues leakage of the bosons and
fermions occurs, and collapse of boson and fermion clouds is
observed. The collapse was found for the following critical
numbers of bosons and fermions: $N_{Bc}\approx 10^{5}; N_K\approx
2\times 10^{4}$. For investigating the stability diagram of the
K-Rb system the mean- field model of Gross-Pitaevskii (GP) type,
coupled to the Thomas-Fermi equation for fermions has been used
\cite{inst3, Roth02, Modugno03}. The ground state and the
stability of the system against collapse was considered with an
imaginary-time evolution scheme. The signature of the instability
is the failure of the convergence procedure toward the ground
state of the system, characterizing by an indefinite growth of the
central density \cite{Modugno03}.

In this paper we study the instability and collapses of the
trapped boson-fermion mixture due to the boson-fermion attractive
interaction, using the effective Hamiltonian for the Bose system
\cite{ChRyz04_PRA,ChRyz04_JETPL}. The effective Hamiltonian
incorporates the three- particle elastic collisions induced by the
boson- fermion interaction. We analyze quantitatively properties
of the $^{87}$Rb and $^{40}$K mixture with an attractive
interaction between bosons and fermions at $T=0$. The stability of
this system on the basis variational condensate wave function was
studied in \cite{ChRyz04_JETPL}, and good agreement with
experiment \cite{sci02bf} was found. We estimate the instability
boson number $N_{Bc}$ for the collapse transition by numerical
calculation of the modified GP equation and give a comparison with
the similar picture in a single Bose- condensate with attractive
interaction. Our instability analysis will involve dependences of
the chemical potential $\mu$ and the number of boson particles $N$
on the value of central density $n_c$ of the bose- condensate wave
function $\phi({\bf r})$. Considerations based on $n_c$-dependence
were introduced earlier in Ref. \cite{Houbiers96} for studying the
stability of a Bose condensate of atomic $^7$Li in a magnetic trap
at nonzero temperature. The calculations \cite{Houbiers96}
confirmed that the gas becomes mechanically unstable when the free
energy of the system as a function of the central density of the
gas approaches a maximum value. In our case we present arguments
based on the second variation of the energy functional. We exhibit
explicitly those variations of the condensate wave function
$\delta\phi({\bf r})$ that reveal an instability point of energy
functional and relate their to a broad class of variations which
does not preserve the normalization.

\section{Effective Bose Hamiltonian}

First of all we briefly discuss the effective boson Hamiltonian
\cite{ChRyz04_PRA,ChRyz04_JETPL}.  Our starting point is the
functional-integral representation of the grand-canonical
partition function of the Bose-Fermi mixture. It has the form
\cite{popov1,stoof1}:
\begin{eqnarray}
Z&=&\int
D[\phi^*]D[\phi]D[\psi^*]D[\psi]\exp\left\{-\frac{1}{\hbar}\left(
S_B(\phi^*,\phi)+\right.\right. \nonumber\\
&+&\left.\left.S_F(\psi^*,\psi)+S_{int}(\phi^*,\phi,\psi^*,\psi)
\right)\right\} \nonumber
\end{eqnarray}
and consists of an integration over a complex field
$\phi(\tau,{\bf r})$, which is periodic on the imaginary-time
interval $[0,\hbar\beta]$, and over the Grassmann field
$\psi(\tau,{\bf r})$, which is antiperiodic on this interval.
Therefore, $\phi(\tau,{\bf r})$ describes the Bose component of
the mixture, whereas $\psi(\tau,{\bf r})$ corresponds to the Fermi
component. The term describing the Bose gas has the form:
\begin{eqnarray}
S_B(\phi^*,\phi)&=&\int_0^{\hbar\beta}d\tau\int d{\bf r} \left\{
\phi^*(\tau,{\bf r}) \left( \hbar \frac{\partial}{\partial
\tau}-\frac{\hbar^2\nabla^2}{2 m_B} + \right.\right.\nonumber\\
&+& \left.\left. V_B({\bf r})- \mu_B \right) \phi(\tau,{\bf r})+
\frac{g_B}{2}|\phi(\tau,{\bf r})|^4 \right\} \nonumber
\end{eqnarray}
Because the Pauli principle forbids $s$-wave scattering between
fermionic atoms in the same hyperfine state, the Fermi-gas term
can be written in the form:
\begin{eqnarray}
S_F(\psi^*,\psi)&=&\int_0^{\hbar\beta}d\tau\int d{\bf
r}\left\{\psi^*(\tau,{\bf r})\left(\hbar \frac{\partial}{\partial
\tau}-\frac{\hbar^2\nabla^2}{2 m_F}+ \right.\right.\nonumber\\
&+&\left.\left.V_F({\bf r}) -\mu_F\right)\psi(\tau,{\bf r})
\right\} \nonumber
\end{eqnarray}
The term describing the interaction between the two components of
the Fermi-Bose mixture is:
\begin{equation}
S_{int}(\phi^*,\phi,\psi^*,\psi)=g_{BF}\int_0^{\hbar\beta}d\tau\int
d{\bf r} |\psi(\tau,{\bf r})|^2|\phi(\tau,{\bf r})|^2, \nonumber
\end{equation}
where $g_B=4\pi \hbar^2a_B/m_B$ and $g_{BF}=2\pi
\hbar^2a_{BF}/m_I$, $m_I=m_B m_F/(m_B+m_F)$, $m_B$ and $m_F$ are
the masses of bosonic and fermionic atoms respectively, $a_B$ and
$a_{BF}$ are the $s$- wave scattering lengths of boson-boson and
boson-fermion interactions. $K$ and $Rb$ atoms in the trap are in
the potentials with an elongated symmetry ($\lambda$- trap
asymmetry parameter)
$$
V_B({\bf r})= \frac{m_B\omega_B^2}{2}(\rho^2+ \lambda z^2), \quad
V_F({\bf r})= \frac{m_F\omega_F^2}{2}(\rho^2+ \lambda z^2)
$$
The trap parameters $\omega_B$ and $\omega_F$ are chosen in such a
way that $m_B\omega_B^2/2= m_F\omega_F^2/2$, so $\omega_F=
\sqrt{m_B/m_F} \omega_B$. Parameters $\mu_B$ and $\mu_F$ are the
chemical potentials for the bose and fermi systems respectively.
The chemical potential of an ideal fermi gas in a trap is $\mu_F=
\hbar\omega_F(6\lambda N_F)^{1/3}$ \cite{Butts97}.

The integral over Fermi fields
\begin{eqnarray}
Z_F&=&\int
D[\psi^*]D[\psi]\exp\left(-\frac{1}{\hbar}\left(S_F(\psi^*,\psi)+
\right.\right.\nonumber\\
&+&\left.\left.S_{int}(\phi^*,\phi,\psi^*,\psi)\right)\right)
\nonumber
\end{eqnarray}
is Gaussian, we can calculate this integral and obtain the
partition function of the Fermi system as a functional of Bose
field $\phi(\tau, {\bf r})$
\begin{equation}
Z_F= \exp\left(-\frac{1}{\hbar}S_{eff}\right), \quad S_{eff}= \int
\limits_0^{\hbar\beta}d\tau \int d{\bf r}f_{eff}(|\phi(\tau,{\bf
r})|) \nonumber
\end{equation}
Using the fact that due to the Pauli principle (quantum pressure)
the radius of the Bose condensate is much less than the radius of
the Fermi cloud $R_F\approx \sqrt{\mu_F/V_0}$, one can use an
expansions in powers of $V_F({\bf r})/\mu_F$ and obtain the
effective Hamiltonian in the form
\begin{eqnarray} \label{he}
H_{eff}[\phi]&=& \int d{\bf
r}\left\{\frac{\hbar^2}{2m_B}|\nabla\phi|^2+ (V_{eff}({\bf
r})-\mu_B)|\phi|^2 + \right. \nonumber \\
&+& \left.\frac{g_{eff}}{2}|\phi|^4+
\frac{g_{eff}^{BF}}{3}|\phi|^6\right\},
\end{eqnarray}
where
$$
V_{eff}= k_0 \frac{m_B \omega_B^2}{2} r^2, {\;} k_0= (1-
\frac{3}{2}\varkappa \mu_F^{1/2} g_{BF}), {\;} \varkappa=
\frac{\sqrt{2}m_F^{3/2}}{3\pi^2\hbar^3}
$$
$$
g_{eff}= g_B- \frac{3}{2}\varkappa \mu_F^{1/2} g_{BF}^2, \quad
g_{eff}^{BF}= \frac{3\varkappa}{8\mu_F^{1/2}}g_{BF}^3
$$

The first three terms in (\ref{he}) have the conventional
Gross-Pitaevskii \cite{pit} form, and the last term is a result of
boson-fermion interaction. The interaction with Fermi gas leads to
modification of the trapping potential. For the attractive
fermion-boson interaction the system should behave as if it was
confined in a magnetic trapping potential with larger frequencies
than the actual ones, in agreement with experiment \cite{sci02bf}.
Boson-fermion interaction also induces the additional attraction
between Bose atoms which does not depend on the sign of $g_{BF}$.
The last term in $H_{eff}$ (\ref{he}) corresponds to the
three-particle \emph{elastic} collisions induced by the
boson-fermion interaction. In contrast with \emph{inelastic}
3-body collisions which result in the recombination and removing
particles from the system \cite{kagan1, kagan2}, this term for
$g_{BF}<0$ leads to increase of the gas density in the center of
the trap in order to lower the total energy.

\section{Numerical procedure}

To simplify the formalism we introduce dimensionless variables for
the spatial coordinate, the energy, and the wave function as
$$
{\bf r}= a_{\perp} {\bf r'}, \quad E= \hbar \omega_{\perp} E',
\quad \phi({\bf r})= \frac{1}{\sqrt{a^3_{\perp}}} \phi'({\bf r})
$$
where the typical length and energy of the harmonic external
potential are $a_{\perp}= \sqrt{\hbar/m_B\omega_{\perp}}$,
$\hbar\omega_{\perp}= \hbar \omega_B$.

Then the effective Hamiltonian takes the form (the primes omitted)
\begin{eqnarray} \label{E}
H_{eff}&=& \int \left \{ \frac{1}{2}|\nabla \phi|^2+
\left(k_0\frac{\rho^2+ \lambda z^2}{2}- \mu_B\right)|\phi|^2+
\right.
\nonumber\\
&+& \left. \frac{u}{4}|\phi|^4+ \frac{v}{6} |\phi|^6 \right \}d^3r
\end{eqnarray}
where we introduced dimensionless parameters $u= 2
g_{eff}/a_{\perp}^3 \hbar\omega_{\perp}$ and $v=
2g_{eff}^{BF}/a_{\perp}^6 \hbar\omega_{\perp}$. The wave function
$\phi$ is normalized to the number of atoms in the condensate
$\int d^3 r |\phi({\bf r})|^2= N$. In the $T \to 0$ limit
considered, $N$ coincides with the total number of bosonic atoms
in the trap. The explicit form of the ground-state wave function
is obtained by minimizing the energy functional. The first order
variation of the energy functional  gives the modified Gross-
Pitaevskii equation
\begin{equation} \label{GP}
\left(-\frac{\nabla^2}{2}+ k_0\frac{\rho^2+ z^2}{2}- \mu_B+
\frac{u}{2}|\phi|^2+ \frac{v}{2}|\phi|^4\right)\phi= 0
\end{equation}
The parameters of the $^{87}$Rb and $^{40}$K mixture with an
attractive interaction between the bosons and the fermions are the
following \cite{sci02bf}: $a_B=5.25\,\, nm, \
a_{BF}=-21.7^{+4.3}_{-4.8}\,\, nm$. The magnetic potential had an
elongated symmetry, with harmonic oscillation frequencies for Rb
atoms $\omega_{\perp}=\omega_B=2\pi\times 215$ Hz and
$\omega_{B,z}=\lambda\omega_B=2\pi\times 16.3$ Hz. At these
parameter values characteristic length $a_{\perp}= 735$ nm,
chemical potential for fermions $\mu_F \approx 31{\,} \hbar
\omega_B$, $\omega_F \approx 1.47 {\,}\omega_B$, $k_0= 1.07$, $u=
0.11$, $v= -0.0003$. Because of a small deviation $k_0$ from unity
we from now on put $k_0= 1$. Note also that we look for the ground
state of Eq. (\ref{GP}), i.e. the function $\phi({\bf r})$ can be
treated as real one.

To clarify the main features of instabilities of the system we
consider isotropic picture when a problem can be considered
effectively as a one-dimensional. The case of non-spherical
symmetry of the trap is recovered at final stage by multiplying
the critical number of bosons $N_c$ by the reverse trap asymmetry
ratio $1/\lambda$.

So we get the  equation ($\mu \equiv \mu_B$):
\begin{equation} \label{GP2}
\Delta \phi= (r^2- 2\mu+ u\phi^2+ v \phi^4)\phi,
\end{equation}
Solutions of Eq. (\ref{GP2}) will be compared to those for the
single component Bose condensate with attractive interactions. As
an example of Bose system with attractive interaction we choose
$^7$Li \cite{Bradley95}. The $s$-wave scattering length is $a=
-27.3{\,} a_0$, where $a_0$ is the Bohr radius. The transverse
frequency is $\omega_{\perp}/2\pi= 163$ Hz, so the corresponding
characteristic length is $a_{\perp}= 2.97 \cdot 10^{-4}$ cm and
$u= 8\pi a/a_{\perp}=-0.012$.

It is convenient to look for the numeric solutions of Eq.
(\ref{GP2}) introducing the new parameter: central density $n_c=
\phi^2(0)$. Numerical integration of Eq. (\ref{GP2}) with boundary
conditions
$$
\phi'(0)= 0, \qquad \phi(r) {\;} \to {\;} 0, \quad r \to \infty
$$
defines the family of solutions $\phi(r, n_c)$ depending on
central density, the chemical potential $\mu(n_c)$ being also a
function of $n_c$. Such an approach when one considers $n_c$ as an
input parameter except for $\mu$ enables to find solutions in the
region of instability and to go continuously  from stable to
unstable branch in the parameter space. This approach differs from
an imaginary-time scheme \cite{Dalfovo96}, where the stability is
indicated by requiring the convergence procedure to the final
value. Solving Eq. (\ref{GP2}), one can easily estimate the
effective energy $E_{eff}$ corresponding to the functional
(\ref{E}) and the ground state energy $E= E_{eff}+ \mu N$, both as
functions of central density $n_c$.

\begin{figure}
\includegraphics[width=8.5cm]{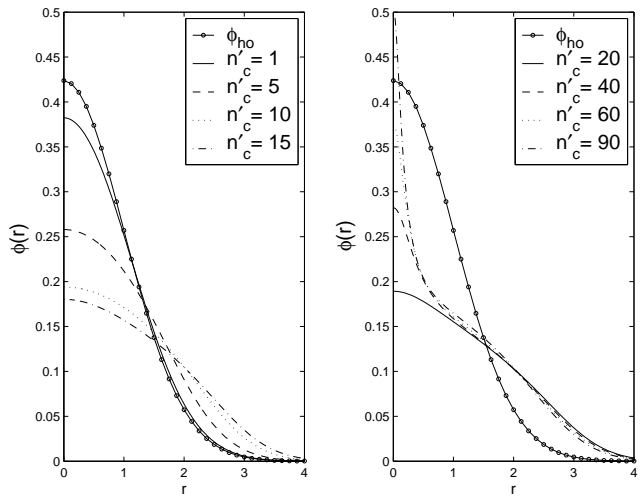}
\caption{\label{BFmix_evol} Evolution of the profile of the
bosonic condensate wave function $\phi(r)$ with increasing the
central density $n'_c$.}
\end{figure}

Derivative $d\phi(r,n_c)/dn_c$ is of especial interest, because it
determines a change in the number of particle through the
variation in the central density $n_c$
$$
\frac{dN}{dn_c}= 2 \int d^3 r {\,} \phi(r,n_c)
\frac{d\phi(r,n_c)}{dn_c}
$$
If at some $n_c$ there comes about $dN/dn_c= 0$ it results in the
appearance of zero mode in density fluctuations and the onset of
instability \cite{Houbiers96}.

\begin{figure}
\includegraphics[width=8.5cm]{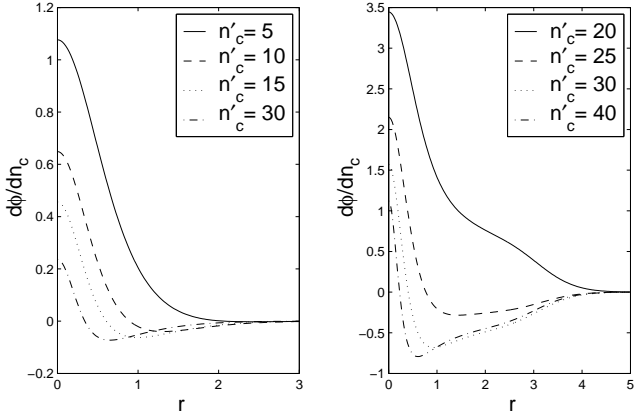}
\caption{\label{Dnc} Evolution of derivative $d\phi(r, n_c)/dn_c$
at various values of central density $n'_c$. Left panel is for the
$^7$Li system, and right panel for BF mixture.}
\end{figure}

It is convenient to consider $\phi(r, n_c)$ and $d \phi(r,
n_c)/dn_c$ as functions of rescaled central density parameter
$n'_c= |u|n_c$. The results are plotted in Fig. \ref{BFmix_evol}
and \ref{Dnc} where $\phi(r, n_c)$ is supposed to be normalized to
1, and distance $r$ is given in units of $a_{\perp}$. Figure
\ref{BFmix_evol} shows the evolution of the profile of the
condensate wave function with increasing the central density. For
comparison, the solution for the isotropic harmonic oscillator,
$\phi_{ho}= \pi^{-3/4}exp(-r^2/2)$, which corresponds to the
ground state of the ideal Bose gas ($u= 0$, $v= 0$), is also
shown. For $n'_c \lesssim 20$, one sees the behavior
characteristic of the Bose gas with repulsion, namely the cloud
density becomes more flat at the trap center, with increasing
radius of the boson cloud. For $n'_c \gtrsim 20$, the solution
changes qualitatively: the central density begins to increase.
Figure \ref{Dnc} shows the evolution of the derivative $d \phi(r,
n_c)/dn_c$ for $^7Li$-system with attractive interaction  ($u <
0$) (the left panel) and for BF mixture (the right panel). The
behavior of BF mixture at relatively high densities ($n'_c \gtrsim
20$) has similar features with $^7Li$-system. Then $n'_c$
increases there is a continues change of the shape of function
$d\phi/dn_c$. It acquires a negative minimum at $r \lesssim
a_{\perp}$, which results in a saturation and a maximum in
$N(n_c)$ dependence.

To relate conditions for stability of a system towards small
changes in its density profile with thermodynamic functions let us
consider the total energy $E$ as a functional of the condensate
wave function $\phi({\bf r})$ and its gradient $\nabla \phi({\bf
r})$
$$
E= \int d^3 r {\,} {\cal E} (\phi({\bf r}), \nabla \phi({\bf r}))
$$
The first order variation $\delta E$ should be considered with the
constraint
\begin{equation} \label{dN}
\delta N= 0, \quad N= \int d^3 r {\,}|\phi({\bf r})|^2
\end{equation}
As usual we broaden the class of allowable variation using the
Lagrange procedure with multiplier $\mu$
$$
E_{eff}= E- \mu N, \quad \delta E_{eff}= 0
$$
Effective energy density contains one more variable $\mu$ ${\cal
E}_{eff}= {\cal E}_{eff}(\phi({\bf r}), \nabla \phi({\bf r}),
\mu)$ and functional $E_{eff}$ coincides with the effective
Hamiltonian (\ref{E}). At $T= 0$ functional E is nothing but the
free energy of the system, and $E_{eff}$ is the thermodynamic
potential $\Omega= F- \mu N$.

Consider functionals $E$ and $E_{eff}$ taking it on the solution
of Eq. (\ref{GP2}) and on particular class of variations
\begin{equation} \label{nc}
\phi(r)= \phi(r,n_c), \quad \delta \phi= \frac{d\phi(r,n_c)}{dn_c}
\end{equation}
Then they become a function of central density $E= E(n_c)$,
$E_{eff}= E_{eff}(n_c)$. The first order variations $\delta
E(\phi, \delta \phi)$ and $\delta E_{eff}(\phi, \delta \phi, \mu)$
considering on functions (\ref{nc}) are nothing but the first
derivatives of functions $E(n_c)$ and $E_{eff}(n_c)$. Due to
equalities
$$
\frac{\delta E}{\delta \phi(r)}= 2\mu\phi, \quad \frac{\delta
E_{eff}}{\delta \phi(r)}= 0, \quad \frac{\partial {\cal
E}_{eff}}{\partial \mu}= -\phi^2
$$
there are simple relations \cite{Houbiers96}
\begin{equation} \label{dEnc}
\frac{dE}{dn_c}= \int d^3 r \frac{\delta E}{\delta \phi(r)}
\frac{d \phi}{d n_c}= \mu(n_c) \frac{dN}{dn_c}
\end{equation}
\begin{equation} \label{dEeffnc}
\frac{dE_{eff}}{dn_c}= \int d^3 r \left\{ \frac{\delta
E_{eff}}{\delta \phi(r)} \frac{d \phi}{d n_c}+ \frac{\partial
{\cal E}_{eff}}{\partial \mu} \frac{d\mu}{dn_c} \right\}=
-\frac{d\mu}{dn_c} N(n_c)
\end{equation}
which relates extremum points of $E(n_c)$ and $E_{eff}(n_c)$ with
an extremum of $\mu(n_c)$ and $N(n_c)$.  Note that variations
(\ref{nc}) do not satisfy constraint (\ref{dN}), which holds only
when $dN/dn_c= 0$. It means that (\ref{nc}) forms a broader class
of variations and include those of them which do not conserve the
number of particle.

Now we relate the behavior of $\mu(n_c)$ and $N(n_c)$ with the
second order variation $\delta^2 E_{eff}(\phi, \delta \phi, \nabla
\delta \phi, \mu)$ taking it on functions (\ref{nc}). It is
related with $d^2 E_{eff}/dn_c^2$ through the equality
\begin{eqnarray}
\frac{d^2E_{eff}}{dn_c^2}&=& \delta^2 E_{eff}(n_c)+ \nonumber \\
&+& \int \left( 2\frac{\partial^2 {\cal E}_{eff}}{\partial \phi
\partial \mu} \frac{d\phi(r,n_c)}{dn_c} \frac{d\mu}{dn_c}+
\frac{\partial {\cal E}_{eff}}{\partial \mu} \frac{d^2\mu}{dn_c^2}
\right) d^3 r \nonumber
\end{eqnarray}
Taking into account that $ \partial^2 {\cal E}_{eff}/\partial \phi
\partial \mu= -2\phi$ we obtain a simple relation
\begin{equation} \label{delta2}
\delta^2 E_{eff}(n_c)= \frac{d\mu(n_c)}{dn_c} \frac{dN}{dn_c}
\end{equation}
Eq. (\ref{delta2}) shows that there is a simple connection between
$\delta^2 E_{eff}(n_c)$ (taken on a particular class of variation)
and the behavior of $\mu(n_c)$ and $N(n_c)$ . At the point of
instability of the system, where $dN/dn_c= 0$, the second
variation $\delta^2 E_{eff}(n_c)$ is equal to zero. As for the
second variation of functional $E$, one can write an equality
$$
\delta^2 E(n_c)= \frac{d\mu(n_c)}{dn_c} \frac{dN}{dn_c}+ 2\mu(n_c)
\int\left(\frac{d\phi(r,n_c)}{dn_c}\right)^2 d^3r
$$
which involves an additional term.

\section{Results and discussion}

\begin{figure}
\includegraphics[width=8.5cm]{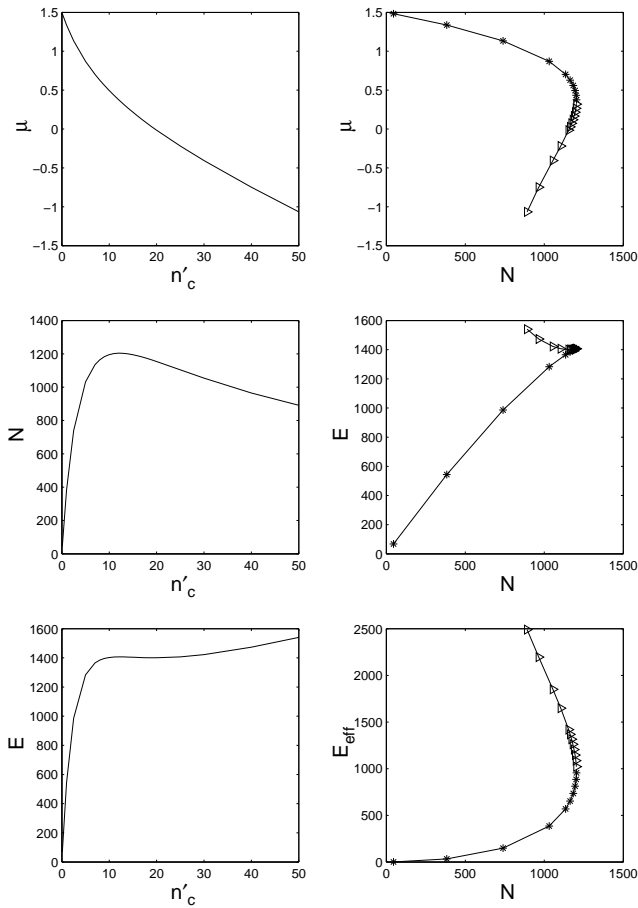}
\caption{\label{attr} $^7$Li system behavior. Left panels:
Chemical potential $\mu$, number of particle $N$ and ground state
energy $E$ as functions of rescaled central density $n'_c$. Right
panels: Chemical potential $\mu$, ground state energy $E$ and
effective ground state energy $E_{eff}$ as functions of number of
particles $N$.}
\end{figure}
\begin{figure}
\includegraphics[width=8.5cm]{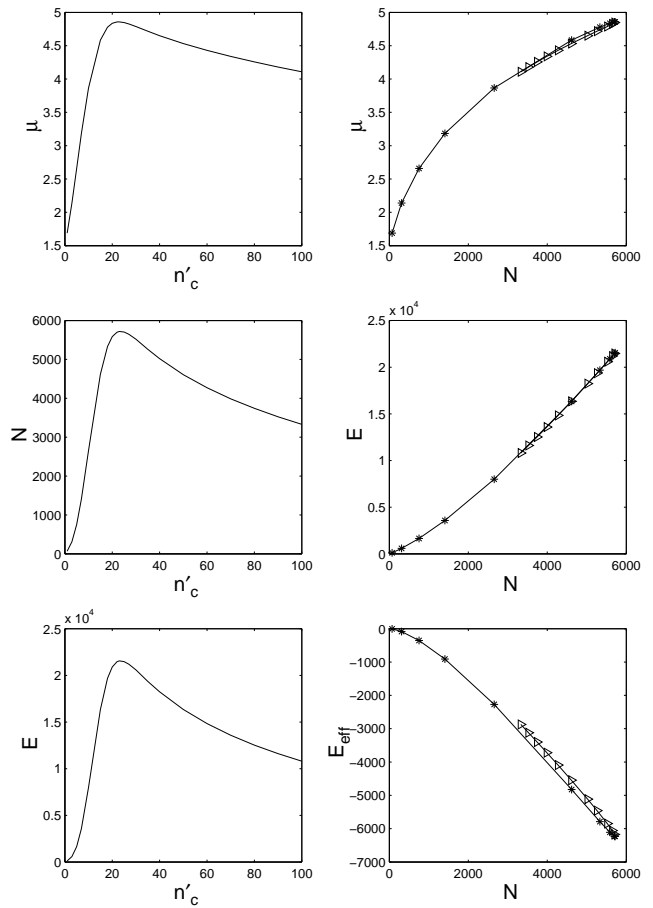}%
\caption{\label{BFmix}BF mixture behavior. The left and right
panels are the same as in Fig. \ref{attr}.}
\end{figure}

To compare the qualitative behavior and properties near collapse
transition of $^7$Li- system with BF mixture, we have calculated
functions $\mu(n'_c)$, $N(n'_c)$ and $E(n'_c)$ shown at left
panels in Fig. 3 and Fig. 4, respectively. Right panels show the
dependencies $\mu$, $E$ and $E_{eff}$ on the number of particle.
Function $\mu(n'_c)$ for $^7$Li system (Fig. 3) gives no sign of
singularities near collapse transition. The same can be said about
$E_{eff}(n'_c)$ due to Eq. (\ref{dEeffnc}). However, there is a
common feature in behavior $^7$Li and BF mixture, namely extremums
of $N(n'_c)$ and $E(n'_c)$. For system with attraction such a
point is at $n'_{c0} \approx 12$, and for BF mixture at $n'_{c0}
\approx 23$. This value of central density corresponds to the
onset of instability of the system towards collapse. This feature
was recognized in Ref. \cite{Houbiers96} and connected with the
presence of zero mode fluctuation of density at this point.

Note, that the extremum for $^7$Li is very wide in $n'_c$. Points
of extremum in $N(n_c)$ and $\mu(n'_c)$ is connected due to Eqs.
(\ref{dEnc}) and (\ref{dEeffnc}) with extremum of $E(n'_c)$ and
$E_{eff}(n'_c)$. Functions $\mu(N)$, $E(N)$, and $E_{eff}(N)$
have forms characteristic of multi-value behavior. Curves with
asterisks are those parts of $\mu$, $E$ and $E_{eff}$ which lie at
$n'_c < n'_{c0}$ and with triangles are those which lie at $n'_c >
n'_{c0}$.

At $n'_c \lesssim 23$ behavior of BF mixture is similar to those
of Bose gas with repulsion. Numerical results show \cite{Ryzhov06,
Ryzhov06_1} that the density profile $n({\bf r})$  can be
accurately described in the framework of Thomas- Fermi (TF)
approximation up to $n'_c \sim 20$. In this region the positive
zero point energy and boson-boson repulsion energy (the first two
terms in Eq. (\ref{he})) stabilize the system. However, if the
central density grows too much, the kinetic energy and boson-boson
repulsion are no longer able to prevent the collapse of the gas.
We see similar behavior of the system with attraction and BF
mixture at $n'_c \gtrsim 23$. Likewise the case of Bose condensate
with attraction (see, for example, \cite{pit,kagan1,kagan2}), the
collapse is expected to occur when the number of particles in the
condensate exceeds the critical value $N_{Bc}$. For BF mixture
curves $\mu(n)$, $E(N)$ and $E_{eff}(N)$ have a point of
termination which corresponds to the maximum number of particle
$N_{cr} \sim 6000$. Taking into account the asymmetry parameter
$\lambda \approx 0.076$ we obtain $N_{cr} \sim 10^5$ which is in
good agreement with the experiment \cite{Bradley95}. A small
difference in $E_{eff}$ for stable and unstable branches arises
solely from a very small difference in chemical potentials of
these states and not connected with computational accuracy. The
small difference in chemical potentials of this branches
($\mu(N)$- curves with asterisks and triangles in Fig. 3) is due
to a small value of the three- body interaction term $v$.

To find the disappearance of the local minimum of functional
$E_{eff}$ which points to the instability of the system, we should
explore the second order variation $\delta^2 E_{eff}$. $\delta^2
E_{eff}$ changes a sign from positive to negative one at the point
of instability. In terms of the steepest descent method absence of
local minimum implies that the convergence towards the local
minimum falls down. The second order variation $\delta^2 E_{eff}$
is given by the quadratic form on $\delta \phi$ and $\nabla \delta
\phi$ and for the functional (\ref{E}) has the form (for our
purposes it is sufficient to consider only real $\delta \phi$)
$$
\delta^2 E_{eff}= \int \left \{ (r^2- 2\mu+ 3u\phi^2+
5v\phi^4)(\delta \phi)^2+ (\nabla \delta \phi)^2 \right \}d^3r
$$
Numerical calculations show that $\delta^2 E_{eff} > 0$ on
solution $\phi= \phi(r,n_c)$ if we take $\delta \phi(r)$ as a
gaussian, satisfying condition (\ref{dN}). This implies that an
extremum of the Hamiltonian is a local minimum. In the case of
Bose condensate with attraction the existence of the barrier
around the metastable state was confirmed in Ref. \cite{Shuryak96}
by extensive variational studies of the nearby wave function.

At the stable branch ($n'_c< n'_{c0}$) the value of $d\mu/dN$ is
negative for $^7Li$ system and is positive for BF mixture. In a
homogeneous one-component system ($N/V= const$) $d\mu/dN$ is
proportional to $\partial \mu/\partial \rho= 1/(\rho^2
\varkappa_T)$ ($\rho= m|\phi|^2$ is the mass density,
$\varkappa_T$ is the isothermal compressibility of the system) and
the criterion of thermodynamic stability $\varkappa_T > 0$ reduces
to the requirement that $d\mu/dN$ should be positive. It is easily
generalized for an inhomogeneous system which can be treated in
the framework of local density approximation. In the local density
approximation the density profile $n(r)=|\phi|^2$ depends on $N$
as a parameter and monotonically expands with increasing of
particle number. So the density $n(r, N)$ undergoes a steady
increase: $dn(r, N)/dN > 0$ at any point $r$ within a radius of
external potential $V_{ext}(r)$. A density profile is determined
from the equation
$$
\mu_{loc}(n(r,N))+ V_{ext}(r)= \mu(N)
$$
So, in the local density approximation criterium of local
stability $\partial \mu_{loc}/\partial n > 0$ through the relation
$d\mu/dN= (\partial \mu_{loc}/\partial n) {\;} (\partial
n/\partial N)$ gives the stability condition $d\mu/dN > 0$ for an
inhomogeneous system in external potential.

\begin{figure}
\includegraphics[width=8.5cm]{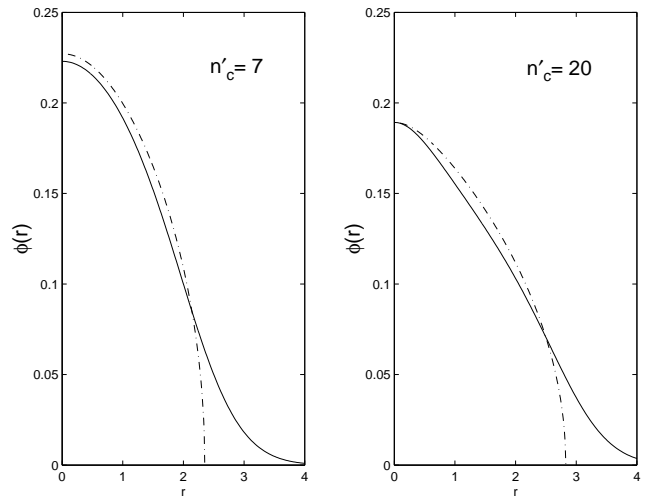}
\caption{\label{BFmix_TF} The profile of the condensate wave
function $\phi(r)$ found from the numeric solution of Eq.
(\ref{GP2}) (solid line) and the TF approximation Eq. (\ref{TF})
(dashed-dotted line).}
\end{figure}

BF mixture at $7 \lesssim n'_c \lesssim 20$ safely can be
considered in the TF approximation  (Fig. \ref{BFmix_TF})
\cite{Ryzhov06_1}. In this case the density profile has the form
\begin{equation} \label{TF}
\phi_{TF}^2(r)= n_{cr} \left[1- \sqrt{1- \frac{R^2-
r^2}{R_{cr}^2}} \right], \quad R^2= 2\mu
\end{equation}
where $R_{cr}^2= u^2/(4|v|)$, $n_{cr}= u/(2|v|)$, $R \le R_{cr}$
and is considered at interval $0 \le r \le R$. The evolution of
the profile corresponds to a monotonic expanding of the boson
cloud with increasing number of bosons. That is why the stable
branch of BF mixture corresponds to the positive value of
$d\mu/dN$.

In contrast, TF approximation is not applicable for $^7Li$ system
for which the sign of $dn(r,N)/dN$ depends on r. At $r \lesssim
a_{\perp}$ one observes a rapid growth of the condensate density,
while at $r > a_{\perp}$ a strong depletion of the condensate
density occurs. The negative value of $d\mu/dN <0$ in this case
can be explained by the following. Using Eq. (\ref{delta2}) the
derivative $d\mu/dN$ in terms of the central density can be
rewritten in the form $d\mu/dN= (d\mu/dn_c) {\;} (dn_c/dN)=
\delta^2 E_{eff}(n_c)/(dN/dn_c)^2$. So the value of $d\mu/dN$
gives, at most, an information about a particular class of
variations (\ref{nc}) of the functional (\ref{E}), which do not
satisfy (\ref{dN}). For variations satisfying condition (\ref{dN})
we have $\delta^2 E_{eff} > 0$. So solution $\phi(r, n_c)$
provides the local minimum of $E_{eff}$ on those functions which
preserve the normalization. Stability of the system is explained
by dynamical reasons, namely for collapse of system to occur,
fluctuations with $\delta N \neq 0$ should exist, which
energetically are not favorable at temperatures under discussion.

In conclusion, our analysis of the stability of K-RB Fermi-Bose
mixture on the basis of effective Bose Hamiltonian shows the clear
resemblance to the behavior of $^7$Li system. There is a value of
central density at which small variations of density profile
conserve the number of particle $\delta N= 0$ and the second
variation $\delta^2 E_{eff}$ changes the sign. The value we
determine for $N_c$ is in very good accordance with experiment.
Points of extremum of functions $\mu(n_c)$ and $N(n_c)$ is related
with the first and the second derivatives of functions $E(n_c)$
and $E_{eff}(n_c)$.

We note that the investigation of the actual dynamics after the
system has been driven into the unstable region would require a
description that go beyond the stationary scenario of Eq.
(\ref{GP2}), in similar fashion to what happens during the
collapse of a single Bose-Einstein condensate with attractive
interaction \cite{kagan1, kagan2}. Here we will not discussed
these aspects, since we are concerned with the determination of
the critical values for the onset of instability. Another
interesting issue concerns the relevance of finite temperature
effects, which are not included in the present treatment.

\section{Acknowledgement}
The work was supported by Russian Science Support Foundation and
the Russian Foundation for Basic Research (Grant No 05-02-17280
and Grant No 04-02-17367).

\end{document}